\documentclass[aps,pra,showpacs,groupedaddress,twocolumn,showkeys]{revtex4-2}
\usepackage{amsmath}
\usepackage{graphicx}
\usepackage{diagbox}
\usepackage{xcolor}
\usepackage[colorlinks=true, letterpaper=true, pdfstartview=FitV, linkcolor=blue, citecolor=blue, urlcolor=blue]{hyperref}

\begin{document}

\title{Transformation of topological optical states via spiral modulation in
fractional-diffraction systems}
\author{Dongdong Wang$^{1,2}$}
\author{Xueqing He$^{1,2}$}
\author{Rujiang Li$^{3}$}
\author{Dumitru Mihalache$^{4}$}
\author{Boris A. Malomed$^{5,6}$}
\author{Pengfei Li$^{1,2}$}
\email{lpf281888@gmail.com}

\affiliation{$^{1}$Department of Physics, Taiyuan Normal University,
Jinzhong 030619, China}
\affiliation{$^{2}$Institute of Computational and
Applied Physics, Taiyuan Normal University, Jinzhong 030619, China}
\affiliation{$^{3}$National Key Laboratory of Radar Detection and Sensing,
School of Electronic Engineering, Xidian University, Xi’an 710071, China}
\affiliation{$^{4}$Horia Hulubei National Institute of Physics and Nuclear
Engineering, Magurele, Bucharest RO-077125, Romania}
\affiliation{$^{5}$Department of Physical Electronics, School of Electrical
Engineering, Faculty of Engineering, and Center for Light-Matter
Interaction, Tel Aviv University, Tel Aviv 69978, Israel}
\affiliation{$^{6}$Instituto de Alta Investigaci\'{o}n, Universidad de
Tarapac\'{a}, Casilla 7D, Arica, Chile}

\begin{abstract}
We propose a scheme for manipulations of a variety of topological states in
fractional optical systems through spiral modulation of the local refraction
index. An analytical approximation, based on a truncated finite-mode system,
and direct simulations reveal that the spiral modulation supports direct
mutual conversion between eigenmodes with topological-charge difference $%
\Delta m=1$, driven by the resonant coupling between the modes and the
underlying spiral modulation. The conversion between eigenmodes with $\Delta
m=2$ requires involvement of an intermediary mode and precise tuning of the
resonance condition. We further explore the conversion of degenerate modes
under the action of azimuthal modulation. Modulated degenerate modes exhibit
incomplete conversion, evolving into intermediate states with odd parity
symmetry. Finally, we examine nonlinear effects on the
spiral-modulation-induced mode conversion, identifying an essential
nonlinearity-induced resonance-frequency shift that critically affects the
conversion efficiency.
\end{abstract}

\maketitle

\section{Introduction}

The fractional Schr\"{o}dinger equation (FSE), as a natural extension of the
conventional Schr\"{o}dinger equation, is the basis of fractional quantum
mechanics~\cite{LASKIN2000298,PhysRevE.62.3135,PhysRevE.66.056108}. The
kinetic-energy operator in FSE is composed of fractional spatial derivatives
with L\'{e}vy index $\alpha $, that usually takes values $1<\alpha \leq 2$,
while conventional quantum mechanics corresponds to $\alpha =2$. The
fractional operator leads to a profound change in the behavior of the wave
function. The experimental realization of FSE requires to design a scheme
for implementing the fractional derivatives. In particular, it was proposed
to do that using one-dimensional L\'{e}vy crystals~\cite{PhysRevE.88.012120}%
. However, building such crystals remains a challenge. Alternatively,
classical optical systems have emerged as a promising platform for the
realization of fractional diffraction. In this connection, it was proposed
by Longhi to implement FSE in optics, designing an aspherical optical cavity
to emulate this equation~\cite{Longhi:15}. This ground-breaking work has
ushered in the realm of the beam dynamics realizing the FSE framework. In
particular, the light-beam propagation in these models gives rise to a
breather-like behavior~\cite%
{PhysRevLett.115.180403,photonics8090353,10.1063/5.0190039}.

A related topic of great interest is the nonlinear FSE, which relies upon
the use of the Kerr nonlinearity of the optical material.\ It produces
diverse families of optical fractional solitons, such as gap solitons~\cite%
{Huang:16}, multipoles~\cite{ZENG2021110589}, trapped modes~\cite{Zeng2020},
Airy waves~\cite{HE2021110470}, as well as solitary vortices~\cite%
{LI2020109783}. Setups supporting fractional solitons offer various
applications for the design of photonic data-processing schemes~\cite%
{doi:10.1126/science.aal5326,Kurtz2020,Helgason2021}. Further, by combining
FSE and a complex potential, the parity-time-symmetry and its breaking may
be featured \cite%
{PhysRevLett.100.103904,Rüter2010,Zhang2016,https://doi.org/10.1002/lpor.201600037,Yao:18}%
. Recently, a research group has used programmable holograms and
single-thermal measurement techniques to simulate L\'{e}vy waveguides and
reconstruct the amplitude and phase of pulses, realizing the fractional Schr%
\"{o}dinger equation in the time domain and observing typical phenomena such
as temporal solitons and pulse splitting due to fractional group-velocity
dispersion~\cite{Liu2023}.

In the field of optics, Rabi oscillations provide the simplest realization
of the resonant mode conversion. In this context, the longitudinally
periodic modulation of the refractive index acts as an effective AC field
which induces the Rabi coupling, resulting in various phenomena, such as the
parametric amplification, inhibition of light tunneling, and the realization
of optical isolation~\cite%
{Kartashov:04,Kartashov:09,Kartashov:14,yu2009complete}. Notably, there
exists a similarity between the stimulated mode-conversion process and Rabi
flopping, which refers to periodic transitions between two stationary states
of a quantum system coupled by an external resonant AC\ field~\cite%
{ROBINETT20041,GARANOVICH20121,https://doi.org/10.1002/lpor.200810055}. In
recent years, Rabi oscillations have been widely investigated in a variety
of optical and photonic systems, including fibers~\cite{Hill:90,Lee:00},
multimode~\cite{PhysRevLett.99.233903,Vysloukh:15,Zhang:15}, coupled \cite%
{Ornigotti_2008} and arrayed~\cite{Makris:08,PhysRevLett.102.123905}
waveguides. two-dimensional (2D) modal structures~\cite%
{doi:10.1126/science.1223824,Kartashov:13}, and waveguides with the
parity-time ($\mathcal{PT}$) or partial $\mathcal{PT}$ symmetry~\cite%
{Vysloukh:14,PhysRevA.109.023515}. Rabi oscillations of topological edge
states and resonant mode conversion in FSE have been studied~too \cite%
{Zhang:17,https://doi.org/10.1002/lpor.201700348,Zhong:19}. Recently, Rabi
oscillations of azimuthons in weakly nonlinear waveguides with a weak
longitudinally periodic modulation of the refractive index and nonlinearity
have been investigated~\cite{10.1117/1.AP.2.4.046002,Wang:23}.

The subject of the present work is resonant mode conversion by spiral
longitudinal modulations in the framework of the nonlinear FSE. First, the
characteristics of eigenmodes in the nonlinear FSE\ are addressed. Then,
resonant mode conversion between modes with different topological charges is
demonstrated under the action of the spiral modulation are demonstrated,
with the help of the resonant coupled-mode theory. The results demonstrate
the conversion between eigenmodes of all orders with topological-charge
differences $1$ and $2$. Additionally, azimuthally-modulated degenerate
superposition modes can be converted into multipole ones of odd orders. The
nonlinearity results in strong asymmetrization of the resonant curves and a
shift of the resonant frequencies.

The paper is organized as follows. In the next section, the model and its
eigenmodes found in the linear regime are presented. Under the action of the
spiral longitudinal modulation, the conversion between the modes with
different topological charges is addressed in Sec.~\ref{sec:linear}. In Sec.
~\ref{sec:nonlinear}, the effect of the cubic nonlinearity on the mode
conversion is studied. The paper is concluded in Sec.~\ref{sec:conclusions}.

\section{The model}

\label{sec:model} We consider the light propagation in a multimode
waveguide, modeled by the scaled nonlinear FSE for the field amplitude $\psi
\left( x,y,z\right) $:
\begin{equation}
i\frac{\partial \psi }{\partial z}=\frac{1}{2}\!\left( -\frac{\partial ^{2}}{%
\partial x^{2}}-\frac{\partial ^{2}}{\partial y^{2}}\right) ^{\alpha
/2}\!\!\!\!\psi -V\!\left( x,y,z\right) \!\psi -\sigma \psi |\psi |^{2}.
\label{main}
\end{equation}%
Here, $x$ and $y$ are the transverse coordinates normalized by the
characteristic scale $r_{0}$, and $z$ is the longitudinal coordinate
normalized by $k_{0}r_{0}^{2}$, where the carrier wavenumber is $k_{0}=2\pi
n_{b}/\lambda _{0}$, with\ wavelength $\lambda _{0}$ in vacuum and
refractive index $n_{b}$. Further, $\alpha $ is the L\'{e}vy index and $%
\sigma $ (that may be positive or negative) is the coefficient of the Kerr
nonlinearity. The spiral waveguide, characterized by strength $p$,
potential rotation pitch $\eta \ll 1$, and spatial spiraling frequency $\Omega$,
is represented by the effective potential
\begin{equation}
V(x,y,z)=p\exp \{-[x+\eta \sin (\Omega z)]^{2}-[y-\eta \cos (\Omega z)]^{2}\}.
\label{V}
\end{equation}

\begin{figure}[h]
\centering\vspace{-0cm} \includegraphics[width=8.5cm]{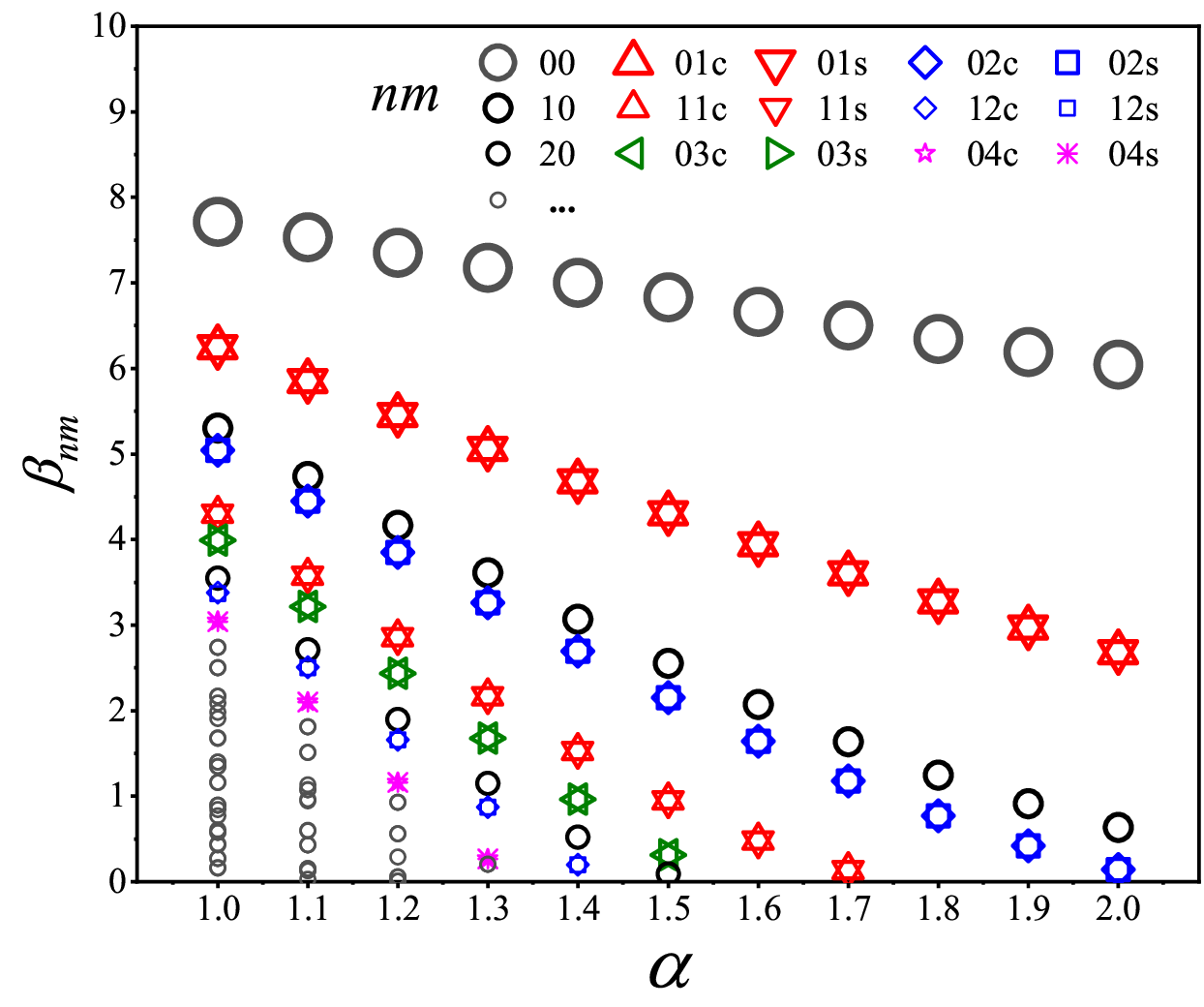} \vspace{-0cm	%
}
\caption{The dependence of eigenvalues $\protect\beta _{nm}$ (propagation
constants), produced by the numerical solution of Eq. (\protect\ref%
{linear_mode_eq}) with $p=10$, on L\'{e}vy index $\protect\alpha $. Symbols $%
\mathrm{c}$ and $\mathrm{s}$, added to mutually degenerate eigenvalues,
indicate their correspondience to the eigenvalues of the cosine and sine
types, respectively.}
\label{fig1}
\end{figure}

First, we consider the linear version of Eq.~\eqref{main} without the
longitudinal modulation, i.e., $\sigma =0$ and $\eta =0$, which supports 2D
linear eigenmodes with propagation constant $\beta $, sought for in the
usual form:
\begin{equation}
\psi (x,y,z)=U(x,y)\exp (i\beta z).  \label{eigenmode}
\end{equation}%
Here linear eigenmode $U(x,y)$ satisfies the stationary equation
\begin{equation}
\beta U=-\frac{1}{2}\left( -\frac{\partial ^{2}}{\partial x^{2}}-\frac{%
\partial ^{2}}{\partial y^{2}}\right) ^{\alpha /2}\!U+pV_{0}\left(
x,y\right) U,  \label{linear_mode_eq}
\end{equation}%
where $V_{0}\left( x,y\right) \equiv \exp (-x^{2}-y^{2})$. Numerical
solution of Eq. (\ref{linear_mode_eq}) by means of the plane-wave expansion
method produces 2D eigenmodes with propagation constant $\beta _{nm}$, as
shown in Fig.~\ref{fig1}. Here, $n$ and $m$ represent the radial node number
and topological charge (vorticity) of these eigenmodes. Note that their
degeneracy is two for $m\geq 1$, corresponding to cosine and sine modes $%
u_{nm}$ and $v_{nm}$, respectively~\cite{PhysRevA.109.023515,Wang:23}.

\begin{figure}[h]
\centering\vspace{-0cm} \includegraphics[width=8.5cm]{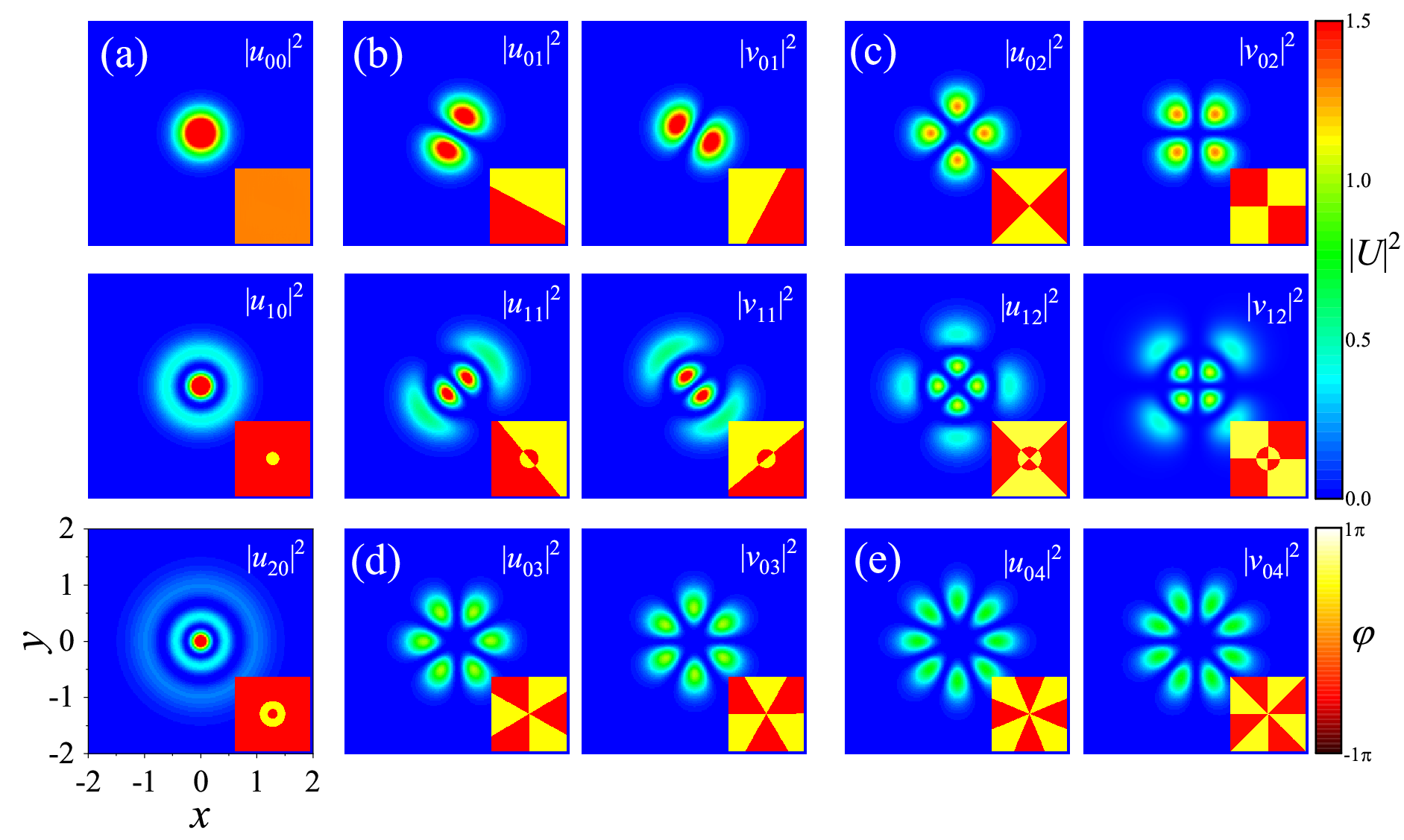} \vspace{-0cm	%
}
\caption{Intensity and phase distributions of the linear eigenmodes for $\protect%
\alpha =1.3$ and $p=10$. (a) The zero-vorticity modes; (b) dipoles; (c)
quadrupoles,;(d) hexapoles; (e) octupoles.}
\label{fig2}
\end{figure}

Further, Fig.~\ref{fig2} displays examples of the intensity distribution in
the first- and higher-order zero-vorticity modes ($u_{00}$, $u_{10}$, and $%
u_{20}$), first- and second-order degenerate dipoles ($u_{01}$, $v_{01}$ and
$u_{11}$, $v_{11}$), first- and second-order degenerate quadrupoles ($u_{02}$%
, $v_{02}$ and $u_{12}$, $v_{12}$), first-order degenerate hexapoles ($%
u_{03} $ and $v_{03}$), and first-order degenerate octupoles ($u_{04}$ and $%
v_{04}$).

\section{The conversion between linear eigenmodes}

\label{sec:linear} In this section, we consider the evolution of eigenmodes
of the linearized version of Eq. (\ref{main}) ($\sigma =0$) in the presence
of spiral modulation ($\eta \neq 0$), and a possibility of the dynamical
conversion between the eigenmodes with different topological charges. To
this end, we consider a superposition of two sets of degenerate eigenmodes:
\begin{align}
\psi (x,y,z)& =\left[ c_{1}(z)u_{nm}(x,y)+c_{2}(z)v_{nm}(x,y)\right]
e^{i\beta _{nm}z}  \notag \\
& \!+\!\left[ c_{1}^{\prime }\!(z)u_{n^{\prime }\!m^{\prime
}}\!(x,y)\!+\!c_{2}^{\prime }\!(z)v_{n^{\prime }\!m^{\prime }}\!(x,y)\right]
e^{i\beta _{n^{\prime }\!m^{\prime }}\!z},  \label{superposition}
\end{align}%
where $\beta _{nm}$ and $\beta _{n^{\prime }m^{\prime }}$ are the
above-mentioned propagation constants corresponding to the degenerate
eigenmodes $u_{nm},v_{nm}$, and $u_{n^{\prime }m^{\prime }}$, $v_{n^{\prime
}m^{\prime }}$, respectively, and weights $|c_{1}(z)|^{2}$, $|c_{2}(z)|^{2}$%
, $|c_{1}^{\prime }(z)|^{2}$, and $|c_{2}^{\prime }(z)|^{2}$ of the modes
and can be calculated as%
\begin{align}
c_{1}(z)& =e^{-i\beta _{nm}z}{\displaystyle\iint }u_{nm}^{\ast }(x,y)\psi
(x,y,z)dxdy,  \notag \\
c_{2}(z)& =e^{-i\beta _{nm}z}{\displaystyle\iint }v_{nm}^{\ast }(x,y)\psi
(x,y,z)dxdy,  \notag \\
c_{1}^{\prime }(z)& =e^{-i\beta _{n^{\prime }m^{\prime }}z}{\displaystyle%
\iint }u_{n^{\prime }m^{\prime }}^{\ast }(x,y)\psi (x,y,z)dxdy,  \notag \\
c_{2}^{\prime }(z)& =e^{-i\beta _{n^{\prime }m^{\prime }}z}{\displaystyle%
\iint }v_{n^{\prime }m^{\prime }}^{\ast }(x,y)\psi (x,y,z)dxdy.  \label{c1c2}
\end{align}

Substituting expression.~\eqref{superposition} in Eq.~\eqref{main} with $%
\sigma =0$, multiplying both sides by $u_{nm}^{\ast }$, $v_{nm}^{\ast }$ and
$u_{n^{\prime }m^{\prime }}^{\ast }$, $v_{n^{\prime }m^{\prime }}^{\ast }$,
respectively, and integrating over the transverse coordinates, we derive the
following system of evolution equations:

\begin{widetext}
\begin{align}
\frac{dc_{1}}{dz}& =ip{\displaystyle\iint }V_{0}G\left[
|u_{nm}|^{2}c_{1}+u_{nm}^{\ast }v_{nm}c_{2}+\left( u_{nm}^{\ast
}u_{n^{\prime }m^{\prime }}c_{1}^{\prime }+u_{nm}^{\ast }v_{n^{\prime
}m^{\prime }}c_{2}^{\prime }\right) e^{-i\left( \beta _{nm}-\beta
_{n^{\prime }m^{\prime }}\right) z}\right] dxdy,  \nonumber \\
\frac{dc_{2}}{dz}& =ip{\displaystyle\iint }V_{0}G\left[ u_{nm}v_{nm}^{\ast
}c_{1}+|v_{nm}|^{2}c_{2}+\left( v_{nm}^{\ast }u_{n^{\prime }m^{\prime
}}c_{1}^{\prime }+v_{nm}^{\ast }v_{n^{\prime }m^{\prime }}c_{2}^{\prime
}\right) e^{-i\left( \beta _{nm}-\beta _{n^{\prime }m^{\prime }}\right) z}%
\right] dxdy,  \nonumber \\
\frac{dc_{1}^{\prime }}{dz}& =ip{\displaystyle\iint }V_{0}G\left[
|u_{n^{\prime }m^{\prime }}|^{2}c_{1}^{\prime }+u_{n^{\prime }m^{\prime
}}^{\ast }v_{n^{\prime }m^{\prime }}c_{2}^{\prime }+\left(
u_{nm}u_{n^{\prime }m^{\prime }}^{\ast }c_{1}+v_{nm}u_{n^{\prime }m^{\prime
}}^{\ast }c_{2}\right) e^{i\left( \beta _{nm}-\beta _{n^{\prime }m^{\prime
}}\right) z}\right] dxdy,  \nonumber \\
\frac{dc_{2}^{\prime }}{dz}& =ip{\displaystyle\iint }V_{0}G\left[
u_{n^{\prime }m^{\prime }}v_{n^{\prime }m^{\prime }}^{\ast }c_{1}^{\prime
}+|v_{n^{\prime }m^{\prime }}|^{2}c_{2}^{\prime }+\left( u_{nm}v_{n^{\prime
}m^{\prime }}^{\ast }c_{1}+v_{nm}v_{n^{\prime }m^{\prime }}^{\ast
}c_{2}\right) e^{i\left( \beta _{nm}-\beta _{n^{\prime }m^{\prime }}\right)
z}\right] dxdy,  \label{weight1}
\end{align}%
\end{widetext}where kernel $G$ may be expanded in powers of the small
spiral-modulation pitch $\eta $:
\begin{align}
G(x,y,z)& =e^{-\eta ^{2}-2\eta \left[ x\sin (\Omega z)-y\cos (\Omega z)\right]
}-1  \notag \\
& =i\eta \left[ \left( x-iy\right) e^{i\Omega z}-\left( x+iy\right)
e^{-i\Omega z}\right]  \notag \\
& -\eta ^{2}-\frac{1}{2}\eta ^{2}\left[ \left( x-iy\right) e^{i\Omega
z}-\left( x+iy\right) e^{-i\Omega z}\right] ^{2}  \notag \\
& +....~.  \label{Expansion}
\end{align}%
Adopting the resonance condition of the coincidence of the spiral-modulation
frequency and a difference between two propagation constants found in the
absence of the modulation,\textit{\ viz}., $\Omega =\beta _{nm}-\beta
_{n^{\prime }m^{\prime }}$, and the rotating-wave approximation (i.e.,
dropping rapidly oscillating terms), we reduce Eqs. (\ref{weight1}) to the
form of
\begin{align}
\frac{dc_{1}}{dz}& =-\eta p\left( D_{1}c_{1}^{\prime }+D_{2}c_{2}^{\prime
}\right) ,  \notag \\
\frac{dc_{2}}{dz}& =-\eta p\left( D_{3}c_{1}^{\prime }+D_{4}c_{2}^{\prime
}\right) ,  \notag \\
\frac{dc_{1}^{\prime }}{dz}& =\eta p\left( D_{1}^{\ast }c_{1}+D_{3}^{\ast
}c_{2}\right) ,  \notag \\
\frac{dc_{2}^{\prime }}{dz}& =\eta p\left( D_{2}^{\ast }c_{1}+D_{4}^{\ast
}c_{2}\right) ,  \label{weight2}
\end{align}%
where $D_{1}={\iint }(x-iy)V_{0}u_{nm}^{\ast }u_{n^{\prime }m^{\prime }}dxdy$%
, $D_{2}={\iint }(x-iy)V_{0}u_{nm}^{\ast }v_{n^{\prime }m^{\prime }}dxdy$, $%
D_{3}={\iint }(x-iy)V_{0}v_{nm}^{\ast }u_{n^{\prime }m^{\prime }}dxdy$, and $%
D_{4}={\iint }(x-iy)V_{0}v_{nm}^{\ast }v_{n^{\prime }m^{\prime }}dxdy$.

Using Eq.~\eqref{weight2}, the frequency of the mode conversion between two
mutually degenerate modes, coupled by the resonant modulation, is
\begin{equation}
\Omega _{R}=2\eta p\left\vert D\right\vert ,  \label{frequency}
\end{equation}%
where $D\equiv D_{1}+D_{2}+D_{3}+D_{4}$ represents the collective coupling
parameter. The mode conversion is governed by phase-matching conditions,
which imply that if and only if the angular-momentum number difference
satisfies condition $\Delta m\equiv |m-m^{\prime }|=1$, the collective
coupling coefficient $D$ attains a nonzero value~\cite{Kartashov:13}, which
is written above. Under this crucial condition, the Rabi frequency $\Omega
_{R}$ determines the efficiency of the inter-mode energy transfer. When $%
\Delta m\neq 1$, the symmetry-enforced destructive interference between the
coupling terms annihilates the collective interaction, thereby completely
suppressing the inter-mode energy conversion.

\begin{figure}[h]
	\centering\vspace{-0cm} \includegraphics[width=8.5cm]{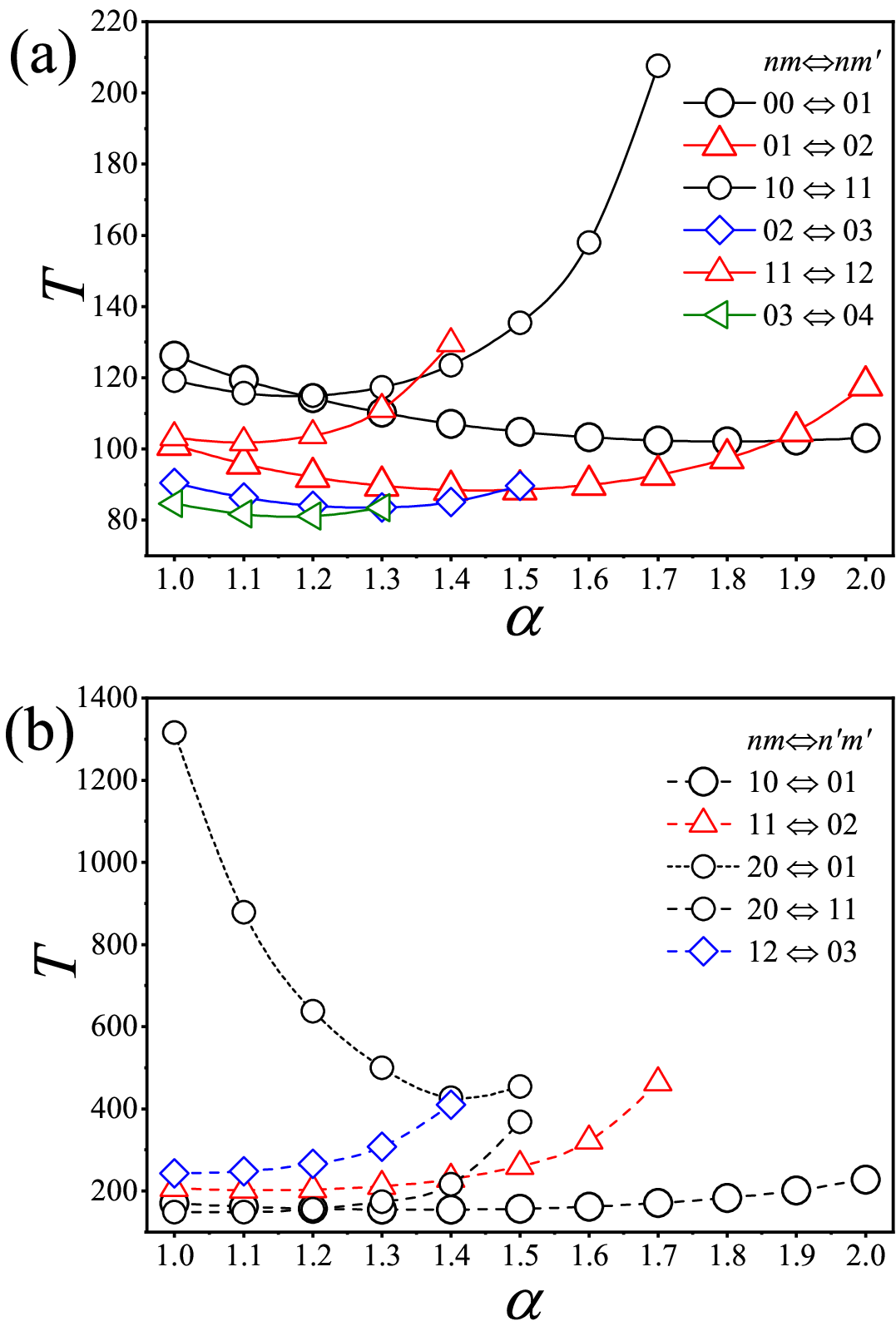} \vspace{-0cm	%
	}
	\caption{Relationship between the conversion period $T$ and the Lévy index $\alpha$. (a) Conversion between modes sharing the same radial index $n$ ($nm \leftrightarrow nm'$). (b) Conversion between modes with distinct radial indices ($nm \leftrightarrow n'm'$).}
	\label{fig3}
\end{figure}

In particular, Eq. \eqref{weight2} predicts the conversion of the
zero-vorticity mode $(m=0)$ to the dipole one $(m^{\prime }=1)$. In this
case, $D_{3}=D_{4}=0$, as the zero-vorticity mode is a non-degenerate one.
The mode conversion can occur between eigenmodes sharing identical radial
node numbers $n$, as well as between those with different values of $n$.

Figure \ref{fig3} shows the dependence of the conversion period $T = 2\pi / \Omega_R$ [where $\Omega_R$ is given by Eq.~(\eqref{frequency})] on  parameter $\alpha$, revealing its crucial role in controlling the rate of the mode energy conversion. Figure \ref{fig3}(a) corresponds to the conversion period between modes with the same radial node number $n$ ($nm \leftrightarrow nm'$), while Fig. \ref{fig3}(b) corresponds to that between modes with different radial node numbers $n$ ($nm \leftrightarrow n'm'$). As $\alpha$ decreases, the waveguide supports an expanded set of eigenmodes with higher radial/azimuthal quantum numbers (Fig. \ref{fig1}), dramatically increasing the number of phase-matched conversion pathways satisfying $\Delta m = 1$. Overall, the conversion period in Fig. \ref{fig3}(b) is significantly larger than that in Fig.~\ref{fig3}(a). Notably, the conversion period $T$ exhibits a non-monotonous trend, first decreasing and then increasing with the decrease of $\alpha$. This phenomenon stems from the dual effect of $\alpha$ on the collective coupling coefficient $D$: on the one hand, a decrease in $\alpha$ leads to more modes, which may enhance the coupling strength for certain mode pairs; on the other hand, very small $\alpha$ weakens mode confinement, causing the mode field distributions to broaden, thereby reducing the overlap integrals and ultimately weakening the coupling and increasing the conversion period $T$. This behavior indicates that, by optimizing $\alpha$, one can find an operating point that provides the minimum conversion period.

\begin{figure}[t]
\centering\vspace{-0cm} \includegraphics[width=8.5cm]{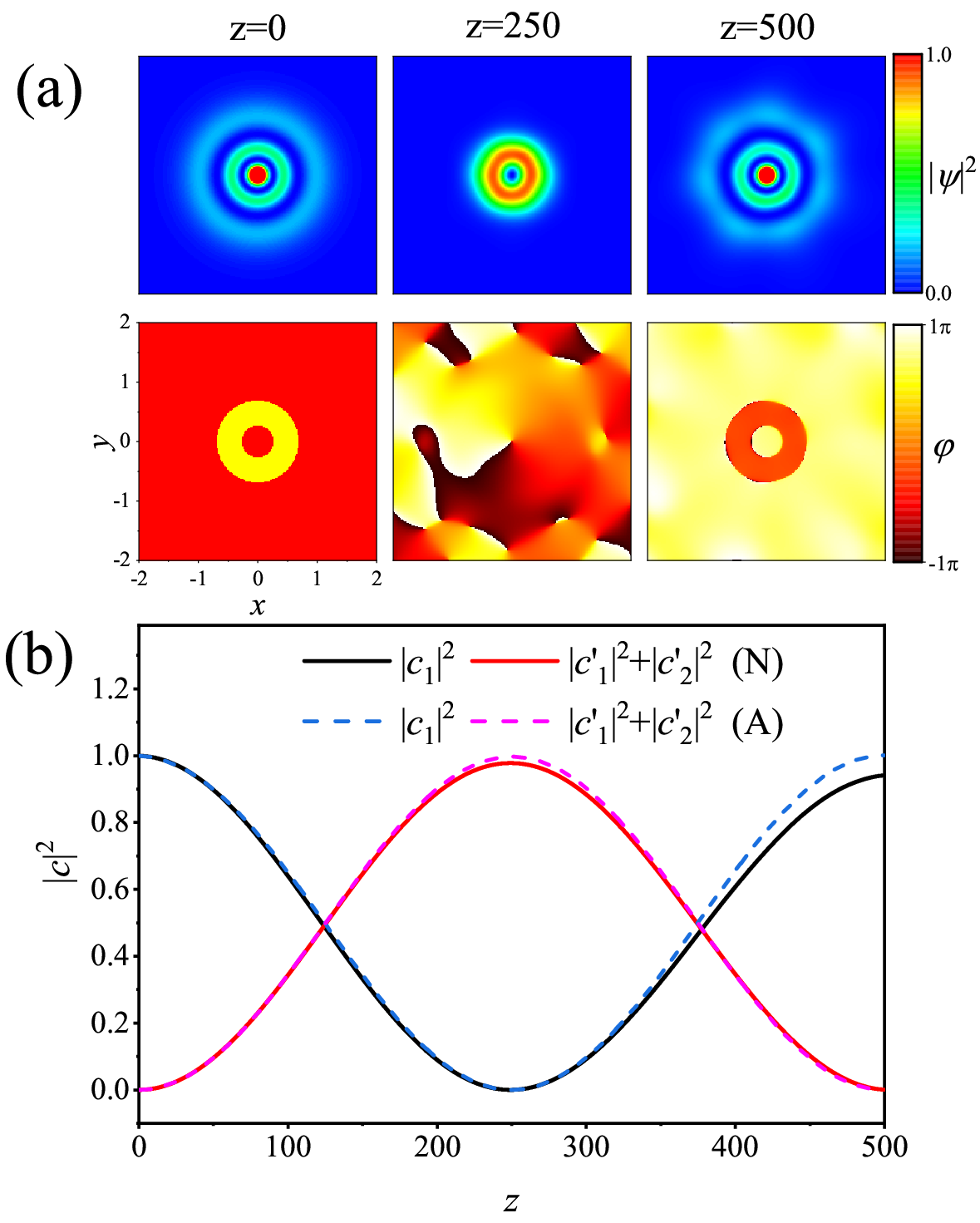} \vspace{-0cm	%
}
\caption{The conversion between the third-order zero-vorticity mode $%
(u_{20}) $ and first-order dipole ones $(u_{01},v_{01})$ at $\protect\alpha %
=1.3$ in the linear regime.(a) Intensity and phase distributions at
different values of the propagation distances and (b) the evolution of the
modes weights with $m=0$ and $m^{\prime }=1$, where the solid and dashed
curves display, respectively, the numerical result and the analytical one
given by Eq.~\eqref{weight2}. The third-order zero-vorticity mode $u_{20}$
was launched into the waveguide at $z=0$. Here, the parameters are $\protect\sigma =0$, $\protect\eta =0.01$, $\protect\beta _{01}=5.0621,\protect\beta %
_{20}=1.1489$, $\Omega =\protect\beta _{01}-\protect\beta _{20}$.}
\label{fig4}
\end{figure}

Figure \ref{fig4} demonstrates the conversion between the third-order zero-vorticity mode $(u_{20})$ and
first-order dipole ones $(u_{01},v_{01})$, with $\Delta m=1$. Under the
action of the spiral modulation, the input mode $u_{20}$ undergoes a
conversion into the degenerate superposed dipole, with the topological
charge $1$, at $z=250$, and then returns to its original mode at $z=500$,
thus completing a cycle of the Rabi oscillations(see Supplemental Material, Animation\_Fig4~\cite{Supplemental}). The numerical result for
the period corroborates the analytical prediction given by Eq.~%
\eqref{frequency}, which yields $T=2\pi /\Omega _{R}=499.7$(Fig.~\ref{fig3}(b)). Note that the
two degenerate dipole modes are always represented by equal weights in the
course of the conversion, i.e., $|c_{1}^{\prime }(z)|^{2}=|c_{2}^{\prime
}(z)|^{2}$.

\begin{figure}[tb]
\centering\vspace{-0cm} \includegraphics[width=8.5cm]{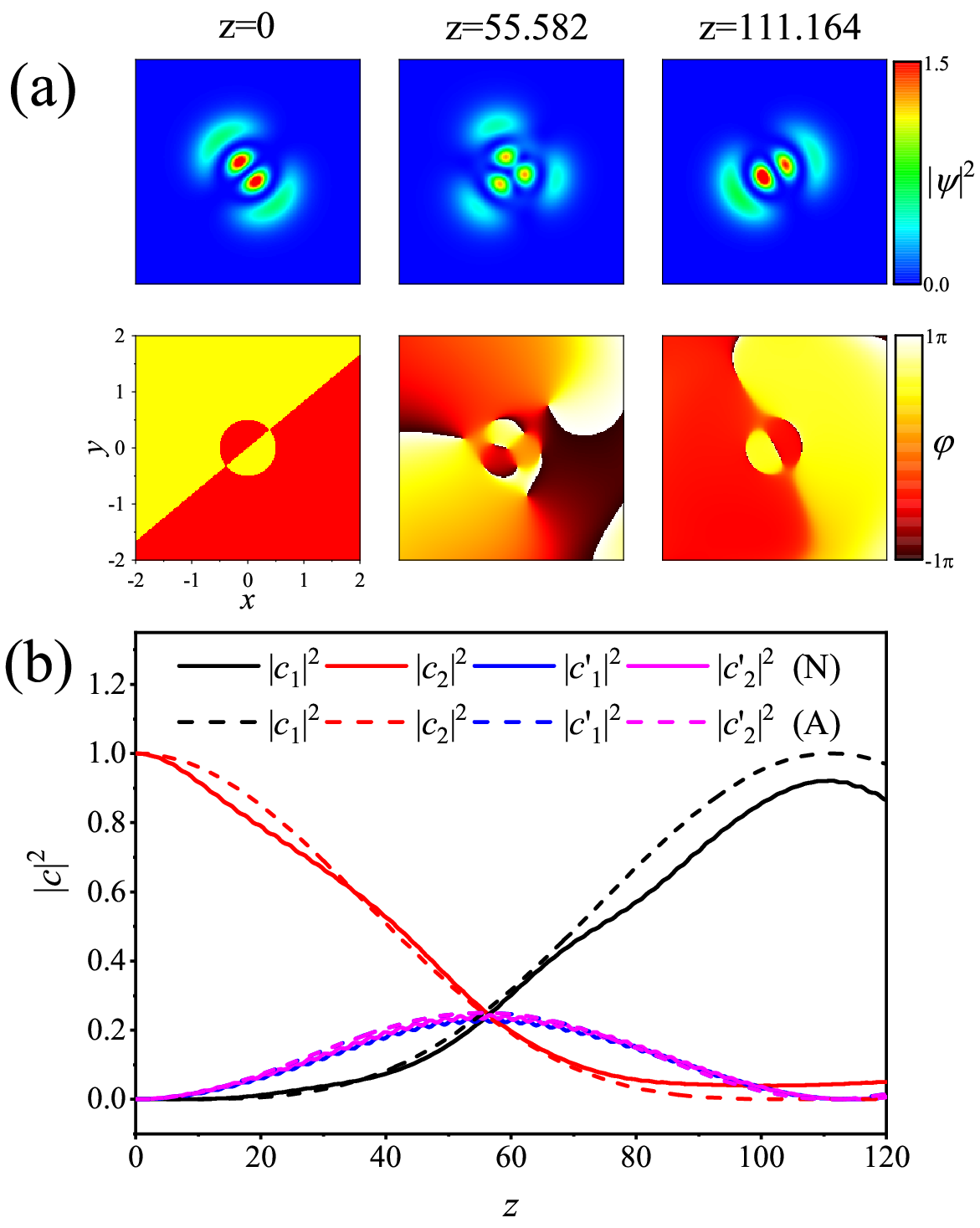} \vspace{-0cm	%
}
\caption{(a) Intensity and phase distributions at different distances, and
(b) the evolution of weights pf the modes with $m=1$ and $m^{\prime }=2$,
where solid and dashed curves represent, respectively, the numerical result
and the analytical one, as produced by Eq.~\eqref{weight2}. The second-order
dipole-sine mode $v_{11}$ was launched into the waveguide at $z=0$. Here,
the parameters are $\protect\alpha=1.3$, $\protect\sigma =0$, $\protect\eta =0.01$, $\protect\sigma %
=0$, $\protect\beta _{11}=2.1684,\protect\beta _{12}=0.8700$, $\Omega =%
\protect\beta _{11}-\protect\beta _{12}$.}
\label{fig5}
\end{figure}

Next we discuss the evolution of\ the input represented by the single
degenerate mode, corresponding to the initial conditions with $c_{2}(0)=1$
and $c_{1}(0)=c_{1}^{\prime }(0)=c_{2}^{\prime }(0)=0$. As shown in Fig.~\ref%
{fig5}, at $z=0$, the input is the second-order dipole mode of the sine
type, $v_{11}$, and we set the resonant frequency $\Omega =\beta _{11}-\beta
_{12}$ to achieve the conversion between two modes of the same order. It is
found that the dipole mode is converted to the tripolar one, with three
peaks (instead of the quadrupole mode) at the half-Rabi-oscillation period, $%
z=55.582$, and returns to the dipole at the full Rabi oscillation period, $%
z=111.164$. However, unlike the original dipole sine mode $v_{11}$, it is
the respective degenerate dipole-cosine mode $u_{11}$(see Supplemental Material, Animation\_Fig5~\cite{Supplemental}). Analyzing the
evolution of the weight curve, it is found that the second-order tripole is
actually formed by a superposition of four components of $u_{11},v_{11}$ and
$u_{12},v_{12}$ with equal weights. During the full Rabi-oscillation period,
$u_{12}$ and $v_{12}$ maintain equal weights, while phase rotation occurs
between the degenerate sine $v_{11}$ and cosine $u_{11}$ modes~\cite{Wang:23}%
. This azimuthal pattern rotation, driven by their interconversion through
the spiral modulation, preserves the dipole configuration while moving its
angular orientation.

\begin{figure}[tb]
\centering\vspace{-0cm} \includegraphics[width=8.5cm]{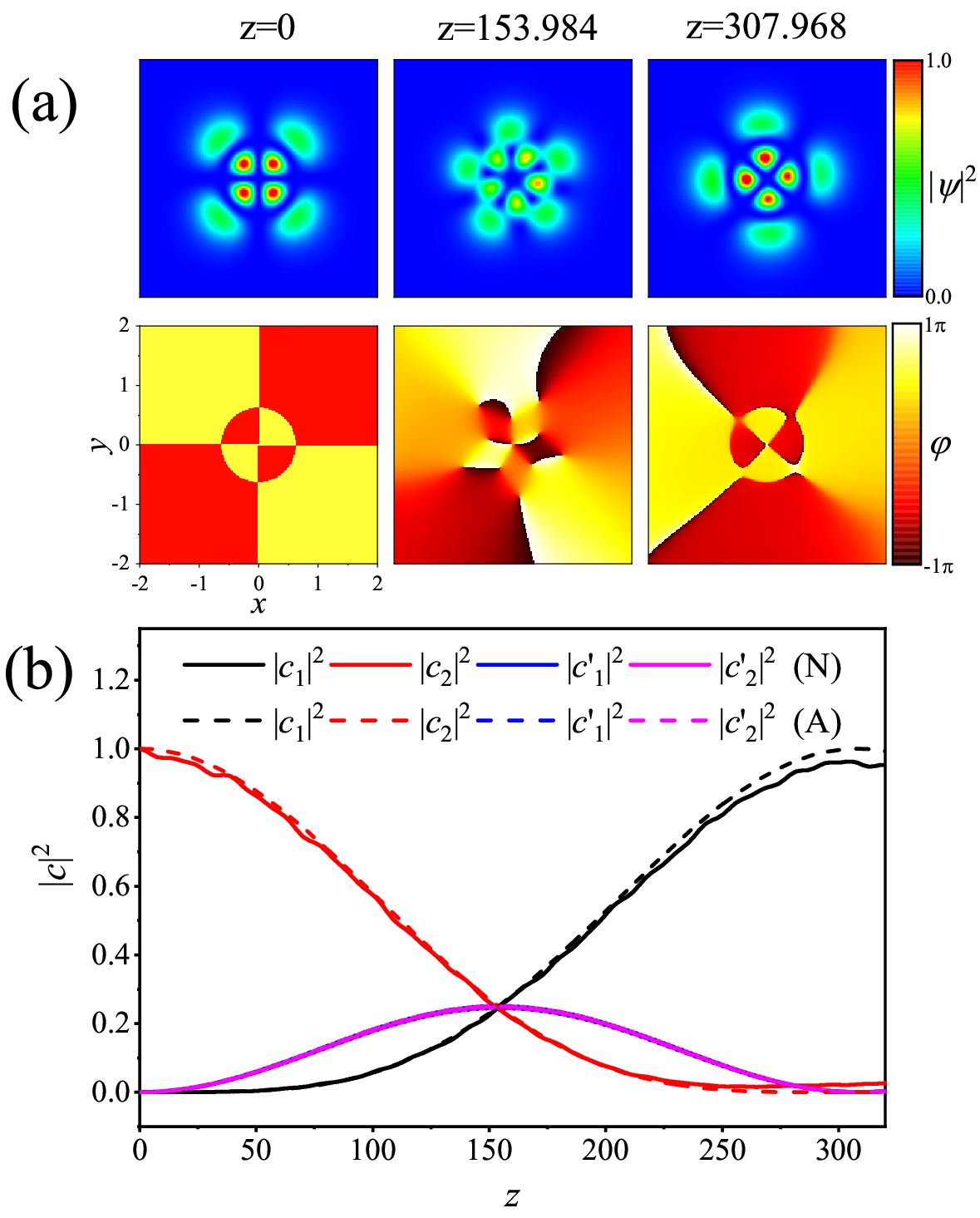} \vspace{-0cm	%
}
\caption{(a) Intensity and phase distributions at different distances and
(b) the evolution of weights of the modes with $m=2$ and $m^{\prime }=3$,
where the solid and dashed curves represent, respectively, the numerical
result and the analytical one, which is produced by Eq.~\eqref{weight2}. The
second-order quadrupole-sine mode $v_{12}$ was launched into the waveguide
at $z=0$. Here, the parameters are $\protect\alpha=1.3$, $\protect\sigma =0$, $\protect\eta =0.01$,
$\protect\beta _{03}=1.6771$, $\protect\beta _{12}=0.8700$, $\Omega =\protect%
\beta _{03}-\protect\beta _{12}$.}
\label{fig6}
\end{figure}

Continuing to consider the conversions between modes of different orders,
Fig.~\ref{fig6} exhibits the evolution of the single second-order quadrupole
mode $v_{12}$(see Supplemental Material, Animation\_Fig6~\cite{Supplemental}), under the resonance condition $\Omega =\beta _{03}-\beta
_{12} $. The conversion creates a pentapole at the half-cycle. It is
constructed as a superposition of two mutually degenerate second-order
quadrupole modes $(u_{12},v_{12})$ and two mutually degenerate first-order
hexapole ones $(u_{03},v_{03})$. In summary, we find that the action of the
external spiral modulation on a single eigenmode sets it in the rotation in
the transverse plane, simultaneously transforming it into an odd-pole
mesostate, with half the conversion efficiency.

To investigate the rotational effect of the spiral modulation, we introduce
azimuthons, which can be constructed from superimposed degenerate linear
eigenmodes ($m\geq 1$) in the framework of the FSE by the following initial
input~\cite{10.1117/1.AP.2.4.046002,Wang:23}:%
\begin{equation}
\psi (x,y,0)=A\left[ u_{nm}(x,y)+iBv_{nm}(x,y)\right] ,  \label{AB}
\end{equation}%
where $A$ is an amplitude factor and $B$ is a parameter which determines the
depth of the azimuthal modulation.

\begin{figure}[t]
\centering\vspace{-0cm} \includegraphics[width=8.5cm]{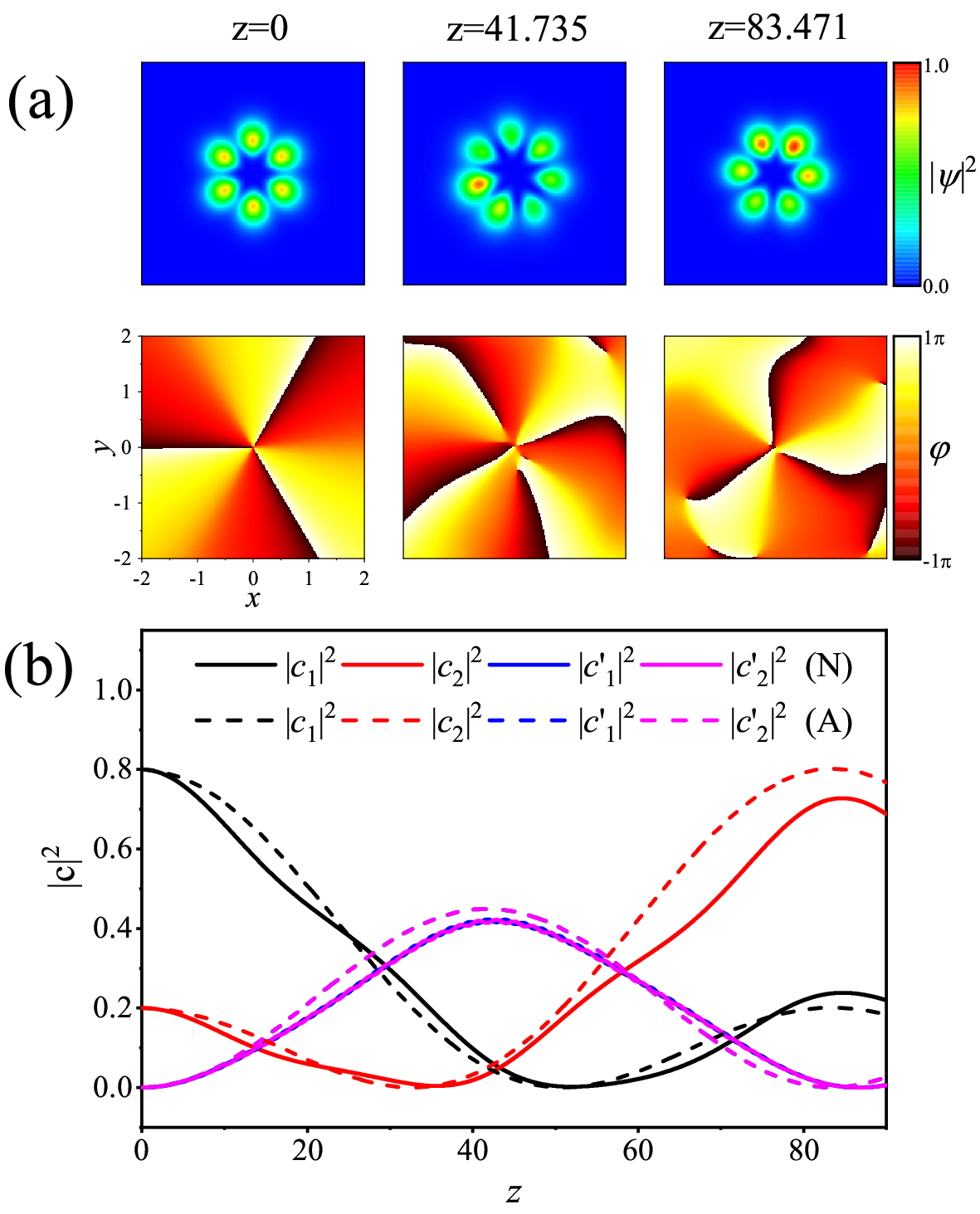} \vspace{-0cm	%
}
\caption{(a) Intensity and phase distributions at different distances and
(b) the evolution of weights of the modes, where the solid and dashed curves
represent, respectively, the numerical result and analytical one produced by
Eq.~ \eqref{weight2}. Here, the parameters are $\protect\alpha=1.3$, $\protect\sigma =0$, $\protect\eta =0.01$, $B=0.5$, $\protect\beta _{03}=1.6771$, $\protect\beta %
_{04}=0.2688$, $\Omega =\protect\beta _{03}-\protect\beta _{04}$.}
\label{fig7}
\end{figure}

Further, in Fig.~\ref{fig7} we use the input in the form of the first-order
hexapole with the azimuthal-modulation intensity of $0.5$ and
spiral-modulation frequency $\Omega =\beta _{03}-\beta _{04}$. In this case,
the sign of the angular momentum of the input azimuthon hexapole agrees with
the direction of the spiral modulation. Therefore, the azimuthon rotates in
the same direction, exhibiting conversion into a heptapole, passing half the
period, $z=41.735$, and returning to hexapole after the full period, $%
z=83.471$(see Supplemental Material, Animation\_Fig7~\cite{Supplemental}). In terms of the weight evolution, the conversion rate is greatly
enhanced by the azimuthal modulation.

The observed minor discrepancy arises from the azimuthal energy
redistribution under the spiral modulation: a non-uniform intensity
distribution between the azimuthal lobes ($\theta $-directional peaks of the
azimuthon) induces energy flow between adjacent lobes. This dynamical
redistribution induces slight radiative losses in the course of the mode
conversion, resulting in a measurable deviation between the predictions of
the coupled-mode theory and numerical simulations.

\begin{figure}[tbh]
\centering\vspace{-0cm} \includegraphics[width=8.5cm]{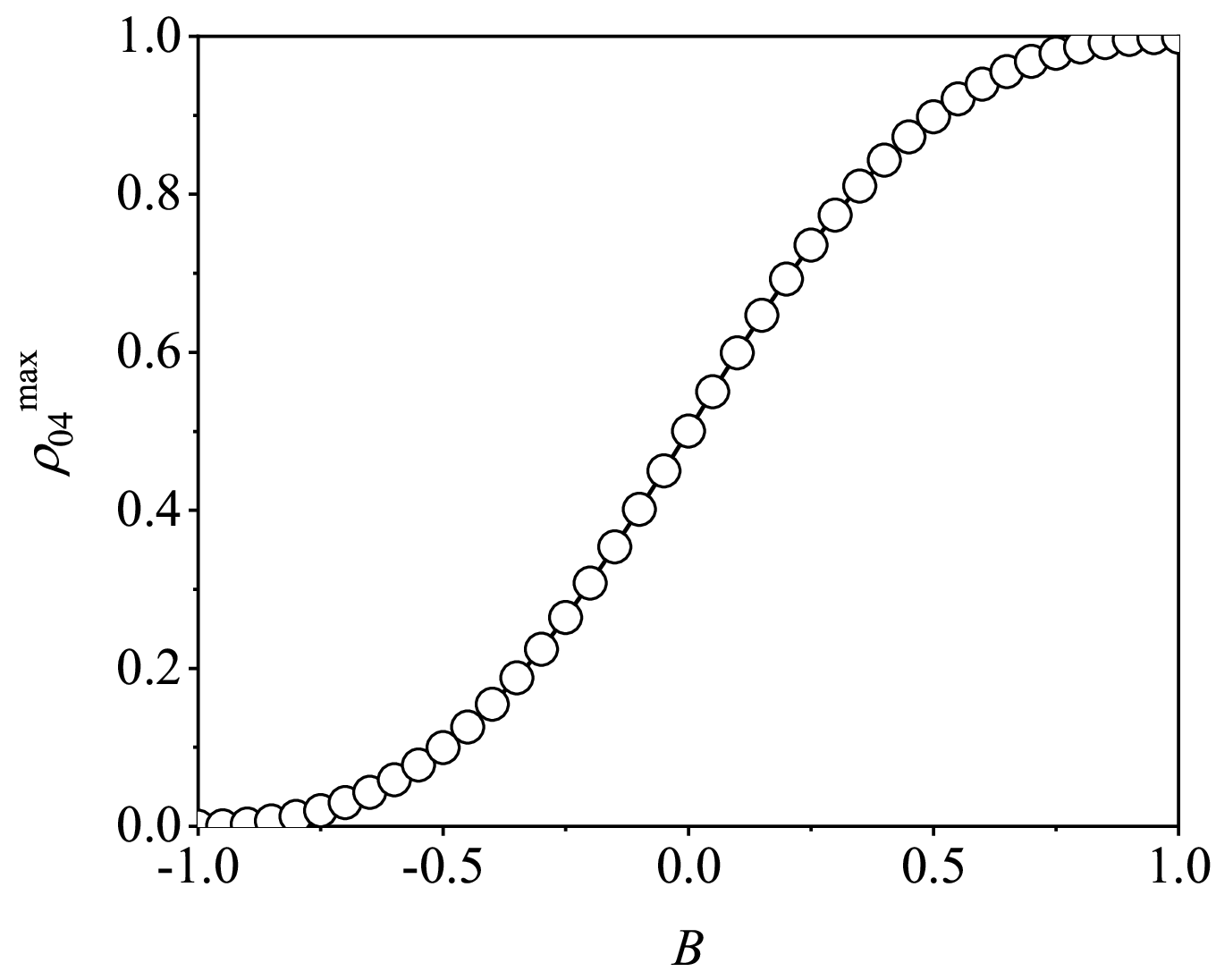} \vspace{-0cm	%
}
\caption{The dependence of the azimuthon's conversion rate on the
azimuthal-modulation depth $B$ in Eq. (\protect\ref{AB}). Here the other
parameters are the same as in Fig~\protect\ref{fig7}.}
\label{fig8}
\end{figure}

Next, we use the coupled-mode theory to explore the dependence between the
azimuthal-modulation depth and the azimuthon's conversion rate. As shown in
Fig.~\ref{fig8}, $\rho _{04}=|c_{1}^{\prime }(z)|^{2}+|c_{2}^{\prime
}(z)|^{2}$ denotes the weight of the octupole $(m=4)$ in the course of the
conversion, and its maximum value represents the conversion rate. When $B>0$
in Eq. (\ref{AB}), the sign of the azimuthon's angular momentum is the same
as the sign of the spiral modulation, and the conversion rate grows with the
increase of the azimuthal-modulation depth. When $B=1$, the equally weighted
degenerate hexapole can be completely converted into the octupole. When $B=0$
, the wave function simplifies to a single constituent of the degenerate
mode pair, achieving a $50\%$ conversion efficiency due to the absence of
the inter-mode coupling. When $B<0$, the sign of the azimuthon's angular
momentum is opposite to that of the direction of the spiral modulation,
resulting in a significant decrease in the conversion rate. The azimuthon
cannot demonstrate complete conversion between modes with $\Delta m=1$ under
the action of spiral modulation. Instead, it is converted into the odd-pole
intermediate state, rotating under the action of the spiral modulation.

Finally, we consider the resonance-driven conversion between eigenmodes with
$\Delta m=2$. To realize this transition, coupling to an intermediate mode
needs to be introduced, therefore we approximate the dynamical state by the
three-component form

\begin{align}
\psi (x,y,z)& =\left[ c_{1}(z)u_{nm}(x,y)+c_{2}(z)v_{nm}(x,y)\right]
e^{i\beta _{nm}z}  \notag \\
& \!+\!\left[ c_{1}^{\prime }\!(z)u_{n^{\prime }\!m^{\prime
}}\!(x,y)\!+\!c_{2}^{\prime }\!(z)v_{n^{\prime }\!m^{\prime }}\!(x,y)\right]
e^{i\beta _{n^{\prime }\!m^{\prime }}\!z}  \notag \\
& +\!\left[ c_{1}^{\prime \prime }(z)u_{n\!^{\prime \prime }m^{\prime \prime
}}\!(x,y)\!+\!c_{2}^{\prime \prime }\!(z)v_{n\!^{\prime \prime }m^{\prime
\prime }}\!(x,y)\right] e^{i\beta _{n^{\prime \prime }m^{\prime \prime }}\!z}
\label{superposition three}
\end{align}%
(cf. Eq. (\ref{superposition}), where $\beta _{nm}>\beta _{n^{\prime
}m^{\prime }}>\beta _{n^{\prime \prime }m^{\prime \prime }}$ are the
propagation constants corresponding to the degenerate eigenmodes $\left(
u_{nm},v_{nm}\right) $, $\left( u_{n^{\prime }m^{\prime }},v_{n^{\prime
}m^{\prime }}\right) $, and $\left( u_{n^{\prime \prime }m^{\prime \prime
}},v_{n^{\prime \prime }m^{\prime \prime }}\right) $, respectively.

Similar to the above consideration, we substituting ansatz~%
\eqref{superposition three} in Eq.~\eqref{main} with $\sigma =0$, multiply
both sides by $u_{nm}^{\ast }$, $v_{nm}^{\ast }$, $u_{n^{\prime }m^{\prime
}}^{\ast }$, $v_{n^{\prime }m^{\prime }}^{\ast }$,$u_{n^{\prime \prime
}m^{\prime \prime }}^{\ast }$ and $v_{n^{\prime \prime }m^{\prime \prime
}}^{\ast }$, respectively, and integrate the result over the transverse
coordinates, omitting rapidly oscillating terms. Thus we derive the
truncated system:

\begin{align}
\frac{dc_{1}}{dz}& =-\eta p\left( D_{1}c_{1}^{\prime }+D_{2}c_{2}^{\prime
}\right) e^{-i\delta z},  \notag \\
\frac{dc_{2}}{dz}& =-\eta p\left( D_{3}c_{1}^{\prime }+D_{4}c_{2}^{\prime
}\right) e^{-i\delta z},  \notag \\
\frac{dc_{1}^{\prime }}{dz}& =\eta p\left( D_{1}^{\ast }c_{1}+D_{3}^{\ast
}c_{2}\right) e^{i\delta z}-\eta p\left( D_{5}c_{1}^{\prime \prime
}+D_{6}c_{2}^{\prime \prime }\right) e^{i\delta z},  \notag \\
\frac{dc_{2}^{\prime }}{dz}& =\eta p\left( D_{2}^{\ast }c_{1}+D_{4}^{\ast
}c_{2}\right) e^{i\delta z}-\eta p\left( D_{7}c_{1}^{\prime \prime
}+D_{8}c_{2}^{\prime \prime }\right) e^{i\delta z},  \notag \\
\frac{dc_{1}^{\prime \prime }}{dz}& =\eta p\left( D_{5}^{\ast }c_{1}^{\prime
}+D_{7}^{\ast }c_{2}^{\prime }\right) e^{-i\delta z},  \notag \\
\frac{dc_{2}^{\prime \prime }}{dz}& =\eta p\left( D_{6}^{\ast }c_{1}^{\prime
}+D_{8}^{\ast }c_{2}^{\prime }\right) e^{-i\delta z},  \label{weight3}
\end{align}%
where $D_{5}={\iint }(x-iy)V_{0}u_{n^{\prime }m^{\prime }}^{\ast
}u_{n^{\prime \prime }m^{\prime \prime }}dxdy$, $D_{6}={\iint }%
(x-iy)V_{0}u_{n^{\prime }m^{\prime }}^{\ast }v_{n^{\prime \prime }m^{\prime
\prime }}dxdy$, $D_{7}\!\!=\!\!{\iint }\!(x-iy)V_{0}v_{n^{\prime }m^{\prime
}}^{\ast }u_{n^{\prime \prime }m^{\prime \prime }}dxdy$, and $D_{8}={\iint }%
(x-iy)V_{0}v_{n^{\prime }m^{\prime }}^{\ast }v_{n^{\prime \prime }m^{\prime
\prime }}dxdy$, $\delta =\beta _{nm}-\beta _{n^{\prime }m^{\prime }}-\Omega
=\Omega -(\beta _{n^{\prime }m^{\prime }}-\beta _{n^{\prime \prime
}m^{\prime \prime }})$. Modes $\left( nm,n^{\prime }m^{\prime }\right) $ and
$\left( n^{\prime }m^{\prime }.n^{\prime \prime }m^{\prime \prime }\right) $
feature the resonance at $(\beta _{nm}-\beta _{n^{\prime }m^{\prime }})$ and
at $(\beta _{n^{\prime }m^{\prime }}-\beta _{n^{\prime \prime }m^{\prime
\prime }})$, respectively. To keep the respective frequency detuning $\delta
$ as low as possible, we set the spiral-modulation spatial frequency to be $%
\Omega =[(\beta _{nm}-\beta _{n^{\prime }m^{\prime }})+(\beta _{n^{\prime
}m^{\prime }}-\beta _{n^{\prime \prime }m^{\prime \prime }})]/2\equiv (\beta
_{nm}-\beta _{n^{\prime \prime }m^{\prime \prime }})/2$. This way, the
spiral modulation drives conversions with $\Delta m=|m-m^{\prime \prime }|=2$%
. Note that, to meet the above conditions, the mode order $n$ should be
specially selected.
\begin{figure}[tbh]
\centering\vspace{-0cm} \includegraphics[width=8.5cm]{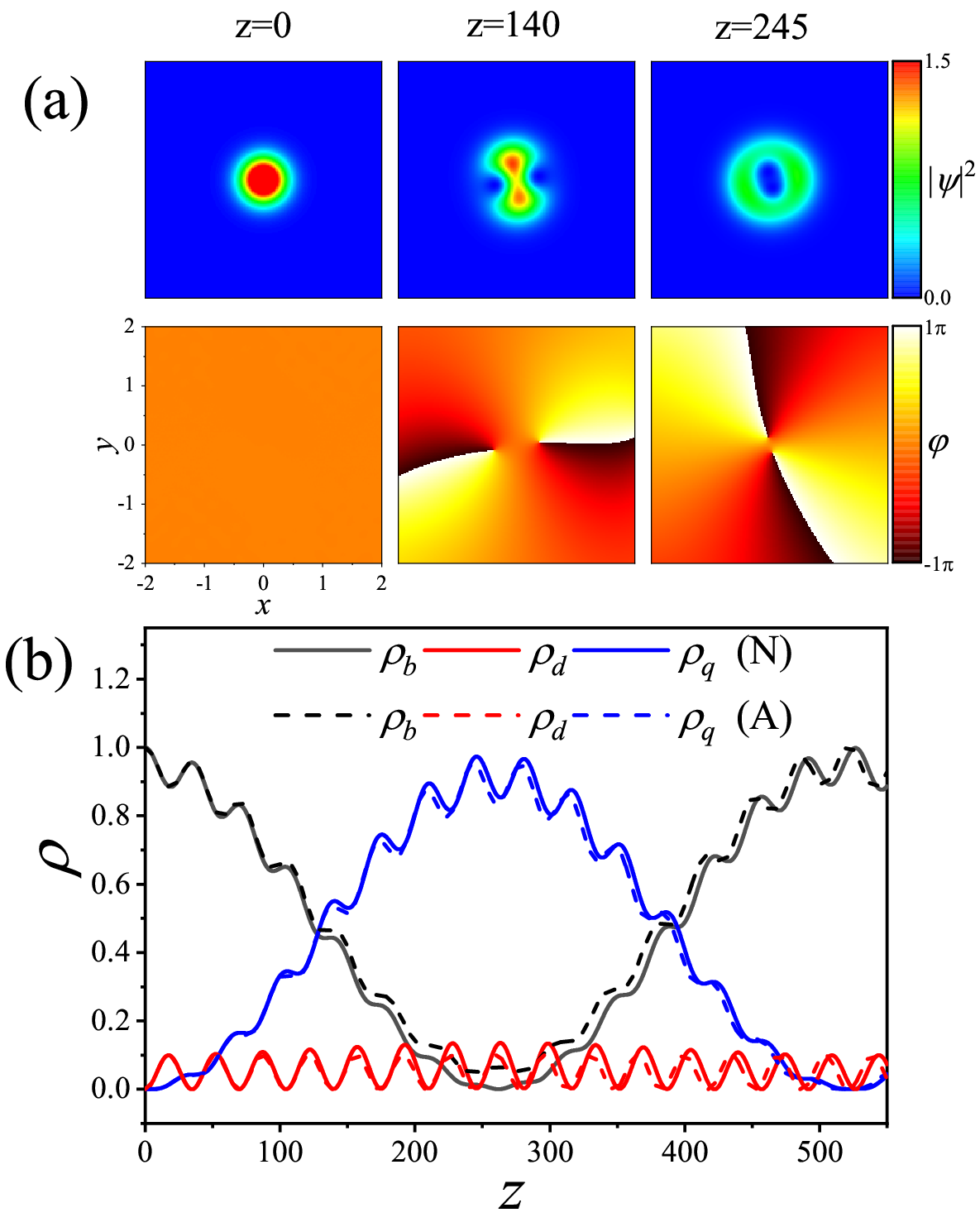} \vspace{-0cm	%
}
\caption{(a) Intensity and phase distributions at different distances and
(b) the evolution of the mode weights with $m=0,m^{\prime }=1$ and $%
m^{\prime \prime }=2$. The solid and dashed curves represent, respectively,
the numerical result and the analytical prediction produced by Eq.~
\eqref{weight3}. The first-order zero-vorticity mode was launched into the
waveguide at $z=0$. The parameters are $\protect\alpha=1.3$, $\protect\sigma =0$, $\protect\eta %
=0.01$ , $\protect\beta _{00}=7.1749$ , $\protect\beta _{01}=5.0621$, $%
\protect\beta _{02}=3.2624$ , $\Omega =\left( \protect\beta _{00}-\protect%
\beta _{02}\right) /2$.}
\label{fig9}
\end{figure}

Figure~\ref{fig9} represents the conversion between the first-order
zero-vorticity and quadrupole modes which $\Delta m=2$, the first-order
dipole mode playing the role of the intermediary. As the zero-vorticity mode
is nondegenerate, the weights of the three modes are $\rho _{b}=|c_{1}|^{2}$%
, $\rho _{d}=|c_{1}^{\prime }|^{2}+|c_{2}^{\prime }|^{2}$ and $\rho
_{q}=|c_{1}^{\prime \prime }|^{2}+|c_{2}^{\prime \prime }|^{2}$,
respectively. As shown in Fig. \ref{fig9}(a), mode $u_{00}$, which is the
input at $z=0$, develops two holes under the action of the spiral
modulation. With the increase of the transmission distance, the holes slowly
move to the center and eventually merge, forming a large hole, so that the
mode evolves into a ring-like structure(see Supplemental Material, Animation\_Fig9~\cite{Supplemental}). As concerns the phase structure,
the two holes represent phase dislocations. In the course of the subsequent
evolution, they fuse into a single phase singularity, which corresponds to
the double topological charge. Eventually, the mode is converted into a
quadrupole one.

Then, from the evolution of the weights observed in Fig.~\ref{fig9}(b), we
conclude that the weight curves exhibit Rabi oscillations. Roughly speaking,
the evolving mode completes the transition from the first-order
zero-vorticity one into the first-order quadrupole and returns to the
zero-vorticity mode after one period of the oscillations. Note that all
degenerate components keep equal weights during the transition. Throughout
the entire process, in addition to the zero-vorticity mode and quadrupole,
components of the dipole mode also appear, and the curves exhibits small
oscillations with the resonance frequency, which causes inherent detuning in
the course of the evolution. The dipole mode, acting as the intermediary,
simultaneously couples to both the zero-vorticity and quadrupole modes,
inducing the generation of the double topological charges and thus enabling
the cross-level transformation.

We stress that the conversion method presented above is different from the
known cascaded scheme for achieving inter-mode conversion~\cite%
{PhysRevLett.99.233903,Wang:23}. In the framework of that scheme, the
conversion starts from the zero-vorticity mode, transforming it into the
dipole one. After completing this stage, the driving frequency is changed to
further convert the dipole mode into the quadrupole one In contrast, the
present conversion method, based in the spiral modulation of the underlying
medium, is different, as the entire process uses the single resonance
modulation frequency. Accordingly, the different conversion techniques
feature different characteristics. For the cascaded scheme, the conversion
efficiency is high, and the stability is strong; however, the required
adjustment method is complex, the admitted bandwidth of the resonance
frequency is extremely narrow, and the resonance conditions are strict,
leading to weak controllability. On the other hand, the spiral-modulation
method, while not enabling complete conversion, requires solely adjusting
the single resonance frequency to directly couple initial and target modes.
By balancing the detuning across the interconnected transitions, this
approach relaxes strict spectral-alignment constraints. Compared to the
cumulative phase-matching limitations of the cascading method, it yields a
broader operational bandwidth and makes the control of the setting
significantly easier.

\section{Effects of nonlinearity on the mode conversion}

\label{sec:nonlinear} In this Section, we aim to demonstrate that efficient
conversion between modes with different topological charges is possible in
the nonlinear medium too, with a resonance shift induced by the
nonlinearity. To address the nonlinear system, we substitute expression~%
\eqref{superposition} in Eq.~\eqref{main} with $\sigma \neq 0$, multiply
both sides by $u_{nm}^{\ast }$, $v_{nm}^{\ast }$, $\ $and $u_{n^{\prime
}m^{\prime }}^{\ast }$, $v_{n^{\prime }m^{\prime }}^{\ast }$, respectively,
and integrate the result over the transverse coordinates, retaining, as in
Eq. (\ref{Expansion}), the first-order expansion term with respect to small
spiral-modulation pitch $\eta $ and dropping rapidly oscillating terms. We
thus derive the following system of nonlinear truncated-mode equations:

\begin{widetext}
\begin{align}
\frac{dc_{1}}{dz}& =-\eta p\left( D_{1}c_{1}^{\prime }+D_{2}c_{2}^{\prime
}\right) e^{i\beta ^{\prime }z}+i\sigma \left\{ Q_{11}\left[ \left(
3\left\vert c_{1}\right\vert ^{2}+2|c_{2}|^{2}\right) c_{1}+c_{1}^{\ast
}c_{2}^{2}\right] +2Q_{12}\rho _{n^{\prime }m^{\prime }}c_{1}\right\} ,
\nonumber \\
\frac{dc_{2}}{dz}& =-\eta p\left( D_{3}c_{1}^{\prime }+D_{4}c_{2}^{\prime
}\right) e^{i\beta ^{\prime }z}+i\sigma \left\{ Q_{11}\left[ \left(
3\left\vert c_{2}\right\vert ^{2}+2|c_{1}|^{2}\right)
c_{2}+c_{1}^{2}c_{2}^{\ast }\right] +2Q_{12}\rho _{n^{\prime }m^{\prime
}}c_{2}\right\} ,  \nonumber \\
\frac{dc_{1}^{\prime }}{dz}& =\eta p\left( D_{1}^{\ast }c_{1}+D_{3}^{\ast
}c_{2}\right) e^{-i\beta ^{\prime }z}+i\sigma \left\{ Q_{22}\left[ \left(
3\left\vert c_{1}^{\prime }\right\vert ^{2}+2|c_{2}^{\prime }|^{2}\right)
c_{1}^{\prime }+c_{1}^{\prime \ast }c_{2}^{\prime 2}\right] +2Q_{12}\rho
_{nm}c_{1}^{\prime }\right\} ,  \nonumber \\
\frac{dc_{2}^{\prime }}{dz}& =\eta p\left( D_{2}^{\ast }c_{1}+D_{4}^{\ast
}c_{2}\right) e^{-i\beta ^{\prime }z}+i\sigma \left\{ Q_{22}\left[ \left(
3\left\vert c_{2}^{\prime }\right\vert ^{2}+2|c_{1}^{\prime }|^{2}\right)
c_{2}^{\prime }+c_{1}^{\prime 2}c_{2}^{\prime \ast }\right] +2Q_{12}\rho
_{nm}c_{2}^{\prime }\right\} ,  \label{nonlinear}
\end{align}
\end{widetext}where $Q_{11}=\frac{1}{3}\iint \left\vert u_{nm}\right\vert
^{4}dxdy=\frac{1}{3}\iint \left\vert v_{nm}\right\vert ^{4}dxdy=\iint
\left\vert u_{nm}\right\vert ^{2}\left\vert v_{nm}\right\vert ^{2}dxdy$, $%
Q_{22}=\frac{1}{3}\iint \left\vert u_{n^{\prime }m^{\prime }}\right\vert
^{4}dxdy=\frac{1}{3}\iint \left\vert v_{n^{\prime }m^{\prime }}\right\vert
^{4}dxdy=\iint \left\vert u_{n^{\prime }m^{\prime }}\right\vert
^{2}\left\vert v_{n^{\prime }m^{\prime }}\right\vert ^{2}dxdy$, $%
Q_{12}=\iint \left\vert u_{nm}\right\vert ^{2}\left\vert u_{n^{\prime
}m^{\prime }}\right\vert ^{2}dxdy=\iint \left\vert u_{nm}\right\vert
^{2}\left\vert v_{n^{\prime }m^{\prime }}\right\vert ^{2}dxdy=\iint
\left\vert v_{nm}\right\vert ^{2}\left\vert u_{n^{\prime }m^{\prime
}}\right\vert ^{2}dxdy=\iint \left\vert v_{nm}\right\vert ^{2}\left\vert
v_{n^{\prime }m^{\prime }}\right\vert ^{2}dxdy$, the respective weights
being $\rho _{nm}=|c_{1}|^{2}+|c_{2}|^{2}$ and $\rho _{n^{\prime }m^{\prime
}}=|c_{1}^{\prime }|^{2}+|c_{2}^{\prime }|^{2}$. Here, $Q_{11}$ and $Q_{22}$
represent the nonlinear self-interaction of the two degenerate modes, while $%
Q_{12}$ governs the cross-interaction between them. The mutual conversion of
the degenerate modes is predominantly driven by the cross-interaction $%
Q_{12} $, whereas the self-interactions $Q_{11}$ and $Q_{22}$ induce the
rotational phase dynamics of the modal profiles~\cite{Wang:23}. As it
follows from Eqs. (\ref{nonlinear}), the nonlinearity break the resonance
condition of the original linear system, making it necessary to introduce a
correction to the resonance modulation frequency. Values of the correction,
which is defined as
\begin{equation}
\beta ^{\prime }=\Omega -(\beta _{nm}-\beta _{n^{\prime }m^{\prime }}),
\label{prime}
\end{equation}%
are displayed below in Fig. \ref{fig10}.
\begin{figure}[tbh]
\centering\vspace{-0cm} \includegraphics[width=8.5cm]{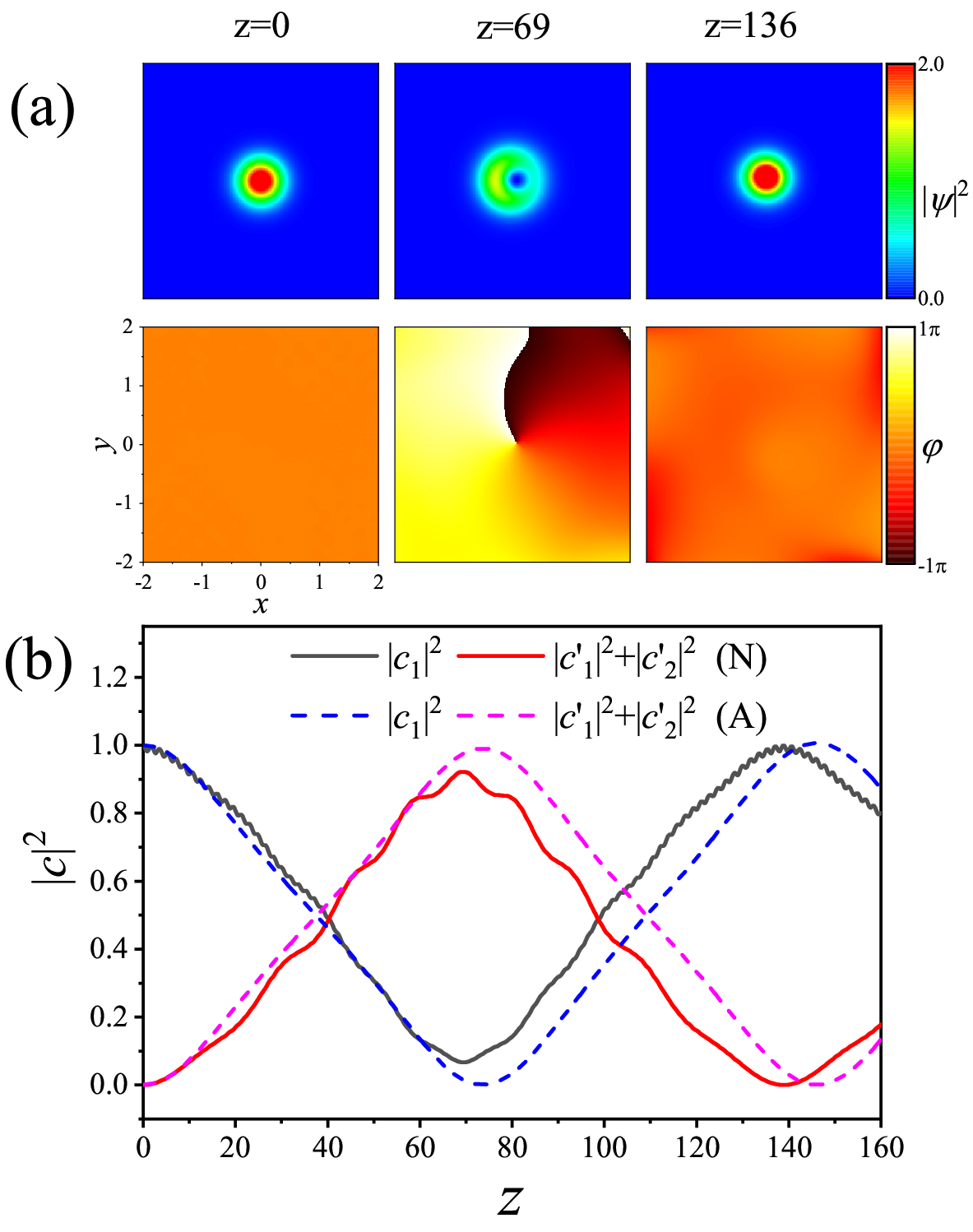} \vspace{-0cm	%
}
\caption{(a) Intensity and phase distributions at different distances and
(b) the evolution of the mode weights in the nonlinear system. The solid and
dashed curves represent, respectively, the numerical results and analytical
predictions produced by Eqs.~\eqref{nonlinear}. The first-order
zero-vorticity mode $u_{00}$ was launched into the waveguide at $z=0$. Here,
the parameters are $\protect\alpha=1.3$, $\protect\sigma =0.3$, $\protect\eta =0.01$, $\protect%
\beta _{00}=7.1749$, $\protect\beta _{01}=5.0621$, $\protect\beta ^{\prime
}=0.09$, $\Omega =\protect\beta _{00}-\protect\beta _{01}+\protect\beta %
^{\prime }$.}
\label{fig10}
\end{figure}
\begin{figure}[tbh]
\centering\vspace{-0cm} \includegraphics[width=8.5cm]{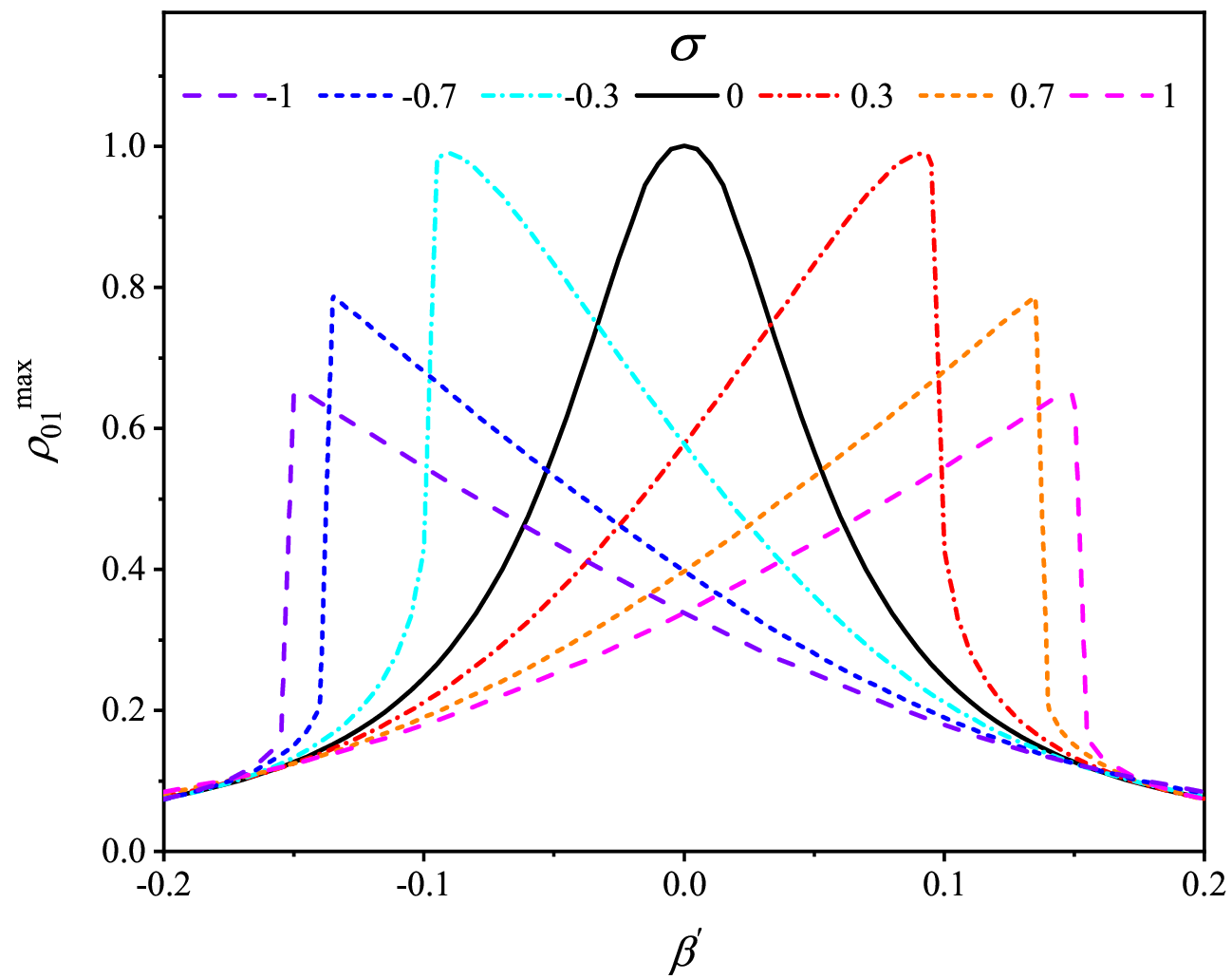} \vspace{%
-0cm	}
\caption{The dependence between the largest value $\protect\rho _{01}^{%
\mathrm{\max }}$ of the mode weight and nonlinear shift $\protect\beta %
^{\prime }$ of the resonance value of the spiral-modulation spatial
frequency for the conversion of the zero-vorticity mode into the dipole one,
see Eq. (\protect\ref{prime}), for different values of the nonlinearity
coefficient $\protect\sigma $ in Eq. (\protect\ref{main}). Other parameters
are the same as in Fig~\protect\ref{fig10}.}
\label{fig11}
\end{figure}

We have also performed direct numerical simulations of the mode evolution in
the framework of the full nonlinear FSE (\ref{main}), as shown, in
particular, in Fig.~\ref{fig10}. The resonance frequency, including the
nonlinearity-induced correction, is given by $\Omega =\beta _{00}-\beta
_{01}+\beta ^{\prime }$, see Eq. (\ref{prime}). Under the action of the
spiral modulation, the zero-vorticity mode gradually transforms into the
dipole one, and then returns to the zero-vorticity mode after the full
cycle(see Supplemental Material, Animation\_Fig10~\cite{Supplemental}). The weight curve demonstrates that, due to the complexity of the
nonlinear system, additional modes are coupled to the mode-transition
process, leading to small oscillations and loss features on the curves,
thereby reducing the conversion rate.

Finally, we use the coupled-mode theory to investigate the
nonlinearity-induced resonance shift. As shown in Fig.~\ref{fig11},
in the absence of the nonlinearity, the resonance curve exhibits the
symmetric peak.
When self-focusing/defocusing nonlinearity is introduced,
with $\sigma >0/\sigma <0$, the peak is deformed in opposite directions. As
the nonlinearity strength increases, the resulting resonance shift becomes
more significant, while the corresponding conversion efficiency decreases
conspicuously.

\section{Conclusion}

\label{sec:conclusions} We have elaborated the scheme for the resonant mode
conversion in the optical waveguide, induced by\ the longitudinal spiral
modulation added to the FSE (fractional Schr\"{o}dinger equation). By adjusting
the L\'{e}vy index $\alpha$, we enable the excitation of higher-order radial modes
and vortex states with extended topological charges which are not stably supported in
conventional nonlinear-Schr\"{o}dinger waveguides, thereby expanding the accessible eigenmode space.
Using the truncated coupled-mode theory, we have predicted that the longitudinal
spiral modulation drives the resonant conversion between
any two eigenmodes with the topological-charge difference $\Delta m=1$, where the
conversion period $T$ can be minimized through the selection
of the optimal value of $\alpha$. The
resonant conversion between two eigenmodes of the same order with $\Delta m=2
$ can be achieved with the help of the intermediary mode. This transition
scheme elaborated here offers greater controllability in comparison to the
known cascading mode-conversion scheme. For degenerate single-component
modes, the longitudinal spiral modulation simultaneously drives both the
rotation of the modes and their conversion, making complete mode conversion
unachievable. However, introducing the azimuthal modulation in the direction
identical to that of the spiral modulation, we can enhance the conversion
efficiency of the azimuthally modulated vortex solitons. The nonlinearity
induces a resonance-point shift, where the self-focusing $(\sigma >0)$ and
defocusing $(\sigma <0)$ nonlinearities give rise to opposite directional
deviations. This shift correlates with the marked reduction in the
conversion efficiency, observed as the nonlinearity strengthens.

\begin{acknowledgments}
	This research was supported by the National National Natural Science Foundation of China (Grants No.11805141); Shanxi Province Basic Research Program (Grants No.202203021222250 and No. 202303021211185); Israel Science Foundation (Grant No.1695/22). R.L. was sponsored by the National Key Research and Development Program of China (Grant No.2022YFA1404900), National Natural Science Foundation of China (Grant No.12104353), and Fundamental Research Funds for the Central Universities (Grant No.QTZX25086).
\end{acknowledgments}

\section*{Data Availability}
The data that support the findings of this article are not publicly available upon publication because it is not technically feasible and/or the cost of preparing, depositing, and hosting the data would be prohibitive within the terms of this research project. The data are available from the authors upon reasonable request.

%\bibliographystyle{plain}
%\bibliography{references_bib}

\end{document}